\documentclass[aps,twocolumn,showpacs]{revtex4}
\usepackage{amsfonts}
\usepackage{graphicx}
\usepackage{dcolumn}
\usepackage{bm,amsmath,amssymb}
\usepackage{epsfig}
\begin{document}

\title{Thermodynamics properties of the dark energy in loop quantum cosmology}
\author{Kui Xiao}
 \email{87xiaokui@mail.bnu.edu.cn}
  \affiliation{Department of Physics, Beijing Normal University, Beijing 100875, China}
\author{Jian-Yang Zhu}
 \email{zhujy@bnu.edu.cn}
  \affiliation{Department of Physics, Beijing Normal University, Beijing 100875, China}
\date{\today}
\begin{abstract}
Considering an arbitrary, varying equation of the state parameter,
the thermodynamic properties of the dark energy fluid in a
semiclassical loop quantum cosmology scenario, which we consider the
inverse volume modification, is studied. The equation of the state
parameters are corrected as a semiclassical one during considering
the effective behavior. Assuming that the apparent horizon has
Hawking temperature, the modified entropy-area relation is obtained,
we find that this relation is different from the one which is
obtained by considering the holonomy correction. Considering the
dark energy is a thermal equilibrium fluid, we get the expressions
for modified temperature, chemical potential and entropy. The
temperature, chemical potential and entropy are well-defined in the
semiclassical regions.
\end{abstract}

\maketitle

\section{Introduction}

More and more evidences \cite{Alessandro-1-0} show that our universe
may be
dominated by dark energy (for more recent review, please read \cite%
{Edmund-1-1} and references therein.). Although the cosmology constant $%
\Lambda$ maybe responsible for the observed data, but it is also the
possible that there are exiting the dynamical mechanism is at work.
There are some candidates to describe those dynamical systems, e.g.,
a canonical scalar field (quintessence) \cite{Guo-1-2}, a scalar
field with negative sign of kinetic term (phantom) \cite{Singh-1-3},
or the combination of quintessence
and phantom in a unified model named quintom \cite{Fang-1-4} and so on \cite%
{Edmund-1-1}. Many authors have analysed the characters of dark energy in
the classical cosmology, e.g., the inflation caused by dark energy \cite%
{Qaisar-1-5}, attractor behavior and solution \cite{Guo-1-6},  the relationship
between geometry and dark energy \cite{Maia-1-8}, the
fate of universe with dark energy \cite{Sami-1-7}, the singularity of
universe with dark energy
\cite{Nojiri-1-9}, the thermodynamical properties of dark energy \cite%
{Pereira-1-11} and so on. And some authors studied the quantum
properties of it \cite{Singh-1-10}. But we still have very little
knowledge about it, and there is not any direct evidence shows
whether the dark energy exists. So it is necessary to study the
properties of it.

The universe can be considered as a thermodynamical system. The
thermodynamical properties of the universe was studied by many authors \cite%
{Hawking-1-13,Wang-1-14}. The apparent horizon and event horizon of
universe are different, the first and second law of thermodynamics
hold on the apparent horizon, but they break down while we consider
the event horizon \cite{Wang-1-14}. According to \cite{Li-1-15}, the
thermodynamical properties of the universe in loop quantum cosmology
(LQC) is still held. But the corrected Firedmann equation will
modify the entropy-area relation of apparent horizon in LQC
\cite{Cai-1-16}. Due to this subtlety, we will focus our attention
on the thermodynamical properties of the apparent horizon first. And
then we will derive thermodynamical variables of dark energy fluid
in the inner side of the apparent horizon in the semiclassical LQC
scenario.

Loop quantum cosmology (for more recent review, please read \cite%
{Bojowald-1-17,Ashtekar-1-18}) is a canonical quantization of
homogeneous spactimes based upon techniques used in loop quantum
gravity (LQG). Due to the homogeneous and isotropic spacetime, the
phase space of LQC is simpler than LQG. The connection is determined
by a single parameter $c$ and the triad is determined by $p$. The
variables $c$ and $p$ are canonically
conjugate with Poisson bracket $\{c,p\}=\frac{8\pi G}{3}\gamma$, in which $%
\gamma$ is the Barbero-Immirzi parameter. In the LQC scenario, the
initial singularity is instead by a bounce. So, thanks for the
quantum effect, the universe is initially contracting phase with the
minimal but not zero volume, and then the quantum effect drives it
to the expanding phase. If one wants to consider the physical effect
near the minimal area, e.g., very small scale factor
$a<a_i=\sqrt{\gamma}l_{\mathrm{pl}}$, it is necessary to consider
the difference equation \cite{Ashtekar-1-19}. But if we consider the
interval $a_i<a<a_\star=\sqrt{\gamma j/3}l_{\mathrm{pl}}$, with the
quantum parameter $j$, we can consider the effective dynamics of
LQC. And the classical cosmology can be recovered if $a>a_\star$. In
this paper, we just consider the thermodynamical of the dark energy
fluid in the semiclassical regions.

In the classical regions, the dark energy thermodynamics has been
studied by many authors \cite{Pereira-1-11}. The thermodynamics
properties of dark
energy fluid described by the equation of state parameter $\omega_{\mathrm{cl%
}}=P_{\mathrm{cl}}/\rho_{\mathrm{cl}}$, in which $P_{\mathrm{cl}},\rho_{%
\mathrm{cl}}$ are the pressure and energy density of dark energy
fluids respective. For the thermodynamical properties of
quintessence filed, with a constant $\omega_{\mathrm{cl}}>-1$, it is
nature to use the results of conventional perfect fluid
thermodynamics \cite{Emmanuel-1-23}. But, in the
case of phantom fields, it is more complex if we consider constant $\omega_{%
\mathrm{cl}}<-1$. Considering the chemical potential $\mu=0$, some
authors showed that the energy density and entropy are positive, but
the temperature is negative \cite{Gonzalez-1-20,Myung-1-21}, and
they argued that this negative temperature due to the quantum
properties of phantom fields. And some other authors showed that, if
the temperature, energy density and
entropy is positive, one must get a negative chemical potential \cite%
{Lima-1-22}. But considering the varying $\omega_{\mathrm{cl}}(a)$, one can
obtain more physics contents, well-defined temperature, energy density,
chemical potential, and entropy on the case of $\omega_{\mathrm{cl}}(a)=-1$,
but the temperature will divergence when $a=0$ in the classical universe \cite%
{Emmanuel-1-23}. In addition, if considering that the temperature of dark
energy fluids have the same temperature of the Hawking temperature on the
apparent horizon, one maybe get negative entropy in condition of $\omega_{%
\mathrm{cl}}(a)<-1$ \cite{Gong-1-24}.

Based on those confused properties of the thermodynamics of dark
energy fluids, it is necessary to discuss it in the semiclassical
regions, even quantum regions. In this paper, we will focus our
interested on the semiclassical one based on the effective LQC. Just
as the above discussion, the scale factor should be in the regions
of $a_i<a<a_\star$. The aim of this paper is tripartite. The first
is to get the modified equation of state parameters
$\omega_{\mathrm{sc}}$ in the semiclassical LQC. The second is to
discuss the modified entropy-area relation of apparent horizon which
is caused by the corrected Firedmann equation. And the last is to
obtain the expressions of thermodynamical variables of dark energy
fluids which is in the inner side of the apparent horizon.

The organization of this paper is as follows. In Sec. \ref{sec2}, we
will introduce the basic concepts of semiclassical LQC and get the
modified varying $\omega_{\mathrm{sc}}(a)$. And then in Section
\ref{sec3}, we will obtain the modified entropy-area relation of the
apparent horizon and the expressions of thermodynamical variables of
the dark energy fluid inside of the apparent horizon. In the last
section \ref{sec4}, we will get some conclusions and discussions.

\section{Modified effective dynamics in LQC}

\label{sec2} The basic variables of phase space of LQC are the
connection $c$ and the triad $p$. In our interested background
spacetime, the connection and triad can be related to the scale
factor, $c=\gamma\dot{a} , p=a^2=V^{2/3}$, with the Barbero-Immirzi
parameter $\gamma\approx0.2375$. The connection and triad satisfy
the Possion bracket
\begin{equation}
\{c,p\}=\frac{8\pi G}{3}\gamma.
\end{equation}
When the scale factor lies in the regions $a_i<a<a_\star$, we
consider the effective dynamics of LQC, with $a_i=\sqrt{\gamma}l_{
\mathrm{pl}},$ $a_\star=\sqrt{\gamma j/3}l_{\mathrm{pl}}$, and $j$
is the quantum parameter and must be taken half integer values. If
$a<a_i$, the quantum effect dominates, we must consider the
difference equation. And  $ a>a_\star$, we turn to the classical
effect. In this paper, we just consider the semiclassical effect of
LQC. The effective Hamiltonian can be written as
\cite{Vandersloot-2-1,Singh-2-1}
\begin{equation}
{\cal{H}}_{\mathrm{eff}}=-\frac{3}{\gamma^28\pi G}\mathcal{S} a
[(c-1)^2+\gamma^2]+{\cal{H}}_{\mathrm{m}}, \label{Heff}
\end{equation}
with the dark energy Hamiltonian ${\cal{H}}_{\mathrm{m}}$. And
$\mathcal{S}$ is
\begin{eqnarray}
{\cal S}
&=&\frac14\left[2((q+1)^3-|q-1|^3)\right.\nonumber\\
&& -3q((q+1)^2-\rm{sgn}(q-1)|q-1|^2)\left.\right]
\end{eqnarray}
with $q=a^2/a_\star^2$.

One can get the equations of motion of connection and triad from the
Hamiltonian (\ref{Heff})
\begin{equation}
\dot{p}=\{p,{\cal {H}}_{{\rm eff}}\}=-\frac{8\pi G\gamma
}3\frac{\partial {\cal {H}}_{{\rm eff}}}{\partial c}=\frac 2\gamma
{\cal S}a(c-1),  \label{dotp}
\end{equation}
\begin{eqnarray}
\dot{c} &=&\{c,{\cal {H}}_{{\rm eff}}\}=\frac{8\pi G\gamma
}3\frac{\partial
{\cal H}_{{\rm eff}}}{\partial p}  \nonumber \\
&=&-\frac{(c-1)^2+\gamma^2}{2\gamma }\left( \frac{{\cal S}}a+\frac{\dot{{\cal S}}}{\dot{a}%
}\right) +\frac{8\pi G\gamma }3\frac{\partial {\cal {H}}_{{\rm
m}}}{\partial p},  \label{dotc}
\end{eqnarray}
where dot denotes the derivation with respect to the time $t$.
Considering Eq.(\ref{dotp}) and remembering the Hamiltonian
constrain satisfies ${\cal{H}}_{\mathrm{eff} }\approx 0$, one can obtain
the modified Friedmann equation
\begin{equation}
H^2=\left(\frac{\dot{a}}{a}\right)^2=\frac{8\pi G} 3\mathcal{S}
\rho_{ \mathrm{sc}} -\frac{\mathcal{S}^2}{a^2},  \label{Fried}
\end{equation}
with the Hubble parameter $H=\dot{a}/a$ and the modified dark energy
density $\rho_{\mathrm{sc}}=E_{\mathrm{m}}/a^3$. $E_{\mathrm{m}}$ is
the eigenvalues for the dark energy Hamiltonian operator that
includes the appropriate modifications to the inverse volume
\cite{Vandersloot-2-1}. We will discuss the definition of
$\rho_{\mathrm{sc}}$ later on.

Considering the modified pressure can be written as
$P_{\mathrm{sc}}=-\frac{
\partial {\cal{H}}_{\mathrm{m}}}{\partial V}=-\frac{2}{3}p^{-1/2}\frac{\partial {\cal{H}}_{
\mathrm{m}}}{\partial p}$, and using the equations of motion of
connection and triad (\ref{dotp}) and (\ref{dotc}), one can obtain
the modified Raychaudhuri equation
\begin{equation}
\frac{\ddot{a}}{a}=-\frac{4 \pi G}{3}\mathcal{S}\left(\rho
_{\mathrm{sc}}+3P_{
\mathrm{sc}}\right)+\frac12 \frac{\dot{\mathcal{S}}}{a}\left(\frac{\dot{a}}{\mathcal{S}}-\frac{\mathcal{S}}{\dot{a}} \right).
\label{Raych}
\end{equation}%
Combining the Friedmann equation (\ref{Fried}) and the Raychaudhuri
equation (\ref {Raych}), one can get
\begin{eqnarray}
\dot{H}&=&\frac{\ddot{a}}{a}-\frac{\dot{a}^{2}}{a^{2}}=-\frac{%
8\pi G}{2}\mathcal{S}(\rho
_{\mathrm{sc}}+P_{\mathrm{sc}})\nonumber\\
&&+
\frac12\frac{\dot{\mathcal{S}}}{a}\left[\frac{\dot{a}}{\mathcal{S}}-\frac{{\mathcal{S}}}{\dot{a}}\right]+\frac{\mathcal{S}^2}{a^2},  \label{dotH}
\end{eqnarray}%
and it is easy to verify the conservation equation
\begin{equation}
\dot{\rho}_{\mathrm{sc}}+3H(\rho _{\mathrm{sc}}+P_{\mathrm{sc}})=0.
\end{equation}%
Defining the equation of state (EoS) parameters $\omega
_{\mathrm{sc}}(a)=P_{ \mathrm{sc}}(a)/\rho _{\mathrm{sc}}(a)$, in
which $\omega _{\mathrm{sc}}(a)$ is a function of scale factor $a$.
We can rewrite the above equation as
\begin{equation}
\dot{\rho}_{\mathrm{sc}}+3\frac{\dot{a}}{a}\rho
_{\mathrm{sc}}[1+\omega _{ \mathrm{sc}}(a)]=0.  \label{Mod-Con-E}
\end{equation}%
Integrating the above equation, it is easy to get the solution for
$\rho _{ \mathrm{sc}}$
\begin{equation}
\rho _{\mathrm{sc}}=\rho _{0}\left\{ \frac{a_{0}^{3[\omega
_{0}+1]}}{ a^{3[\omega _{\mathrm{sc}}(a)+1]}}\right\} \exp \left[
-3\int_{a}^{a_{0}}da \omega _{\mathrm{sc}}^{\prime }(a)\ln a\right]
,  \label{rs}
\end{equation}%
with $\rho _{0},a_{0},\omega _{0}$ denote today's energy density,
scale factor and EoS parameter respectively. For $a_{0}\gg a_{\star
} $, it's not necessary to consider the quantum effect of those
present values (For conservation, we will define the subscript '0'
denotes the present value of a physical quantity.). And the prime
means the derivation with respect to the scale factor $a$.

In the above discussion, we defined the semiclassical energy density
as $ \rho _{\mathrm{sc}}=E_{\mathrm{m}}/a^{3}$, but this is not the
only definition. Singh has showed us that there are two ways to
obtain the semiclassical density \cite{Singh-2-2}. One way is just
like the definition we discussed above. The other one is to define a
density operator $\hat{\rho}
_{\mathrm{q}}=\widehat{{H}_{\mathrm{m}}/a^{3}}$ and then take their
eigenvalues. The eigenvalues for $\hat{\rho}_{\mathrm{q}}$ can be
obtained by considering $\hat{{H}}_{\mathrm{m}}$ and
$\widehat{1/a^{3}}$ \cite {Singh-2-2}
\begin{equation}
\rho _{\mathrm{q}}=d_{j,l}(a)E_{m}(a,\phi
)=D_{l}(q)a^{-3}E_{\mathrm{m} }(a,\phi )=D_{l}(q)\rho
_{\mathrm{sc}},  \label{rs-rq}
\end{equation}%
in which $d_{j,l}=D_{l}(q)a^{-3}$ is the eigenvalue for
$\widehat{1/a^{3}}$ for the large $j$. And $D_{l}$ is given as
\cite{Singh-2-2}
\begin{eqnarray}
D_l(q) &=&\left\{ \frac{27|q|^{1-2l/3}}{8l}\left[\frac{1}{l+3}\left((q+1)^{2(l+3)/3}\right.\right.\right.\nonumber\\
&&\left.-|q-1|^{2(l+3)/3}\right)-\frac{2q}{2l+3}\left((q+1)^{2(l+3)/3}\right.\nonumber\\
&&\left.\left. -\rm{sgn}(q-1)|q-1|^{2(l+3)/3}\right)\right\} ^{\frac 3{2-2l}}, \label{Dl}
\end{eqnarray}
where $q=a^{2}/a_{\star }^{2}$, and $l$ is the quantum ambiguity
parameters with $0<l<1$ \cite{Vandersloot-2-1}. For $a\lesssim
a_{\star }$, $ D_{l}\lesssim 1$, this means $\rho
_{\mathrm{sc}}\lesssim \rho _{\mathrm{q}}$ .

Now, we turn to discuss the relationship between the semiclassical
energy density and the classical one.  The
classical energy conservation is
\begin{equation}
\dot{\rho}_{\mathrm{cl}}+3H(\rho _{\mathrm{cl}}+P_{\mathrm{cl} })=0.
\end{equation}%
Considering the varying EoS parameter $\omega _{\mathrm{cl}}(a)=\rho
_{ \mathrm{cl}}/P_{\mathrm{cl}}$, the above equation can be written
as
\begin{equation}
\dot{\rho}_{\mathrm{cl}}+3H\rho _{\mathrm{cl}}[1+\omega _{
\mathrm{cl}}(a)]=0.  \label{Con-E}
\end{equation}
It is easy to obtain the expression for $\rho _{\mathrm{cl}}$ from the above
equation
\begin{equation}
\rho _{\mathrm{cl}}=\rho _{0}\left\{ \frac{a_{0}^{3[\omega
_{0}+1]}}{ a^{3[\omega _{\mathrm{cl}}(a)+1]}}\right\} \exp \left[
-3\int_{a}^{a_{0}}da \omega _{\mathrm{cl}}^{\prime }(a)\ln a\right]
.  \label{rc}
\end{equation}
Considering the definition of $\rho _{\mathrm{q}}$, we can get the
eigenvalues for $\widehat{\rho}_{\mathrm{q}}$ using the eigenvalues
for $\widehat{1/a^{3}}$, $ d_{j,l}$, instead of $a^{-3}$ in
Eq.(\ref{rc}). So, the semiclassical expression for $\rho
_{\mathrm{q}}$ can be written as
\begin{eqnarray}
\rho _{\mathrm{q}} &=&\rho _{0}\left\{ {a_{0}^{3[\omega _{0}+1]}}{
d_{j,l}^{[\omega _{\mathrm{cl}}(a)+1]}}\right\} \exp \left[
-3\int_{a}^{a_{0}}da\omega _{\mathrm{cl}}^{\prime }(a)\ln a\right]
\notag
\\
&=&D_{{l}}^{\omega _{\mathrm{cl}}(a)+1}\rho _{\mathrm{cl}}(a).
\label{rq}
\end{eqnarray}%
Remembering the relationship between $\rho _{\mathrm{q}}$ and
$\rho_\mathrm{sc}$, which is described by Eq.(\ref{rs-rq}), we can
get the relationship between $\rho _{\mathrm{sc}}$ and $\rho
_{\mathrm{cl}}$
\begin{equation}
\rho _{\mathrm{sc}}=D_{{l}}^{\omega _{\mathrm{cl}}(a)}\rho
_{\mathrm{ cl}}.\label{rsc-rcl}
\end{equation}%
Now we want to get the relationship between $\omega
_{\mathrm{sc}}(a)$ and $ \omega _{\mathrm{cl}}(a)$. Differentiating
$\ln \rho _{\mathrm{sc} }$ with respect to $\ln a$, one can obtain
\begin{equation}
\frac{d\ln \rho _{\mathrm{sc}}}{d\ln a}=\ln D_{l}\frac{d\omega
_{\mathrm{cl} }(a)}{d\ln a}+\omega _{\mathrm{cl}}(a)\frac{d\ln
D_{l}}{d\ln a}+\frac{d\ln \rho _{\mathrm{cl}}(a)}{d\ln a}.
\end{equation}
Note that $\frac{d\ln \rho _{\mathrm{sc}}(a)}{d\ln a}=\frac{d\rho
_{\mathrm{ sc}}(a)}{H\rho _{\mathrm{sc}}(a)dt}=-3[1+\omega
_{\mathrm{sc}}(a)]$, and $ \frac{d\ln \rho _{\mathrm{cl}}(a)}{d\ln
a}=\frac{d\rho _{\mathrm{cl}}(a)}{ H\rho
_{\mathrm{cl}}(a)dt}=-3[1+\omega _{\mathrm{cl}}(a)]$, it is easy to
get
\begin{equation}
\omega _{\mathrm{sc}}(a)=\omega _{\mathrm{cl}}(a)\left[ 1-\frac{1}{3}\ln
D_{l}\frac{d\ln |\omega _{\mathrm{cl}}(a)|}{d\ln a}-\frac{1}{3}\frac{d\ln
D_{l}}{d\ln a}\right].  \label{os-oc}
\end{equation}%
Remembering that $\omega _{\mathrm{cl}}(a)$ may be a negative value
for dark energy fluid, so we need choose the absolute value when we
calculate its logarithm, but it dose not change the answer. Considering Eq.(\ref{os-oc}), it is easy to find that the effective EoS in loop quantum cosmology is possible "-1" crossing. If $\omega_{\mathrm{cl}}=\mathrm{const.}$, Eq.(\ref{os-oc}) will still hold, but the second term in the square brackets is zero. So the constant classical EoS will be a varying term in the effective region.

In this section, we get the equation of the effective EoS of dark energy with a varying EoS in loop quantum cosmology. Now, we can turn to consider the thermodynamical properties of dark energy in loop quantum cosmology scenario.

\section{Thermodynamics properties of dark energy}

\label{sec3} In this section, we will discuss the thermodynamics
properties of dark energy in semiclassical regions. This section
includes two subsections. In the first subsection, we will discuss
the entropy-area relation of apparent horizon in semiclassical
scenario. In the next subsection, we will show the temperature,
chemical potential and entropy of dark energy fluid inner of the
apparent horizon.

\subsection{Modified entropy-area relation}

According to the fact that the Einstein equation can be rewritten as
the form $ dE_{\mathrm{m}}=TdS+W_{\mathrm{m}}dV$ \cite
{Frolov-3-1,Danielsson-3-2,Bousso-3-3}, with the total energy
$E_{\mathrm{m} }=\rho V$ and the work density
$W_{\mathrm{m}}=(\rho-p)/2$, the authors of \cite{Cai-1-16} showed
that the entropy-area relationship of apparent horizon would be
corrected by the modified Friedmann equation in loop quantum cosmology. In this subsection, we
will discuss the modified entropy-area relation which is caused by
the corrected Friedmann equation (\ref{Fried}). Just as
\cite{Cai-1-16} stated, the temperature of apparent horizon is just
determined by the spacetime metric, so it is convenient to assume
the temperature of apparent horizon is $T_{\mathrm{A}}={1}/({2\pi
\tilde{r}_{\mathrm{A}}})$, with the radius of the apparent horizon
$\tilde{r}_{\mathrm{A}}$. Noticed that the LQC is just correcting
the Hamiltonian, not the Lagrangian. And the spacetime metric is
described by the Lagrangian, so it is natural to believe that the
definition of temperature of the apparent horizon in LQC is as same
as in the classical one.

Our interested spacetime is the closed Friedmann-Robertson-Walker
(FRW) universe which is described  by the metric
\begin{eqnarray}
ds^2&=&-dt^2+a^2\left(\frac{dr^2}{1-r^2}+r^2d\Omega_2^2\right)\nonumber\\
&=&h_{ab}dx^adx^b+\tilde{r}^2d\Omega_2^2,  \label{FRW}
\end{eqnarray}
with $h_{ab}=\mathrm{diag}(-1,a^2/(1-r^2)),  \tilde{r}=a(t)r$, and $k$ denotes the spatial curvature. If we define
the Missner-Sharpe mass  \cite{Gong-1-24,Poisson-3-5,Misner-3-6}
\begin{equation}
h^{ab}\partial_a\tilde{r}\partial_b\tilde{r}\equiv f ,  \label{M-S mass}
\end{equation}
the apparent horizon is  determined from $f=0$. So, we can get the radius of
the apparent  horizon
\begin{equation}
\tilde{r}_{\mathrm{A}}=\frac{1}{\sqrt{H^2+1/a^2}}.
\end{equation}
with the Hubble parameter $H$ described by Eq.(\ref{Fried}).

In this paper, we consider the dark energy fluid is a perfect fluid
satisfies
$T_{\mu\nu}=(\rho_{\mathrm{sc}}+P_{\mathrm{sc}})U_{\mu}U_{\nu}+P_{
\mathrm{sc}} g_{\mu\nu}$, where  $\rho_{\mathrm{sc}}$ and
$P_{\mathrm{sc}}$ are the modified energy density and pressure in
the  semiclassical LQC, respectively. And the energy conservation
law which  is described by Eq.(\ref{Mod-Con-E}) still satisfied.

The Einstein equation is written as
\begin{equation}
dE=A\Psi+WdV,  \label{EinEq}
\end{equation}
where $A=4\pi\tilde{r}_{\mathrm{A}}^2$ is the area of the apparent
horizon and $V=\frac{ 4\pi}{3}\tilde{r}_{\mathrm{A}}^3$ is the
volume of 3-dimensional sphere.  $E=M$ is the Misser-Sharp energy
which is given by Eq.(\ref{M-S mass}). $W$ is the work density and
$\Psi$ is the energy-supply  vector, are defined as
\begin{eqnarray}
W=-\frac12T^{ab}h_{ab}, \\
\Psi_{a}=T^{b}_{a}\partial_b\tilde{r}+W\partial_a\tilde{r},
\end{eqnarray}
where $T_{ab}$ is the projection of the $(3+1)$-dimensional
energy-momentum tensor $T_{\mu\nu}$ of dark energy fluid in the FRW
universe in the normal direction of 2-sphere\cite{Cai-1-16}.
Considering our interested metric (\ref{FRW}), we can get the
obvious  expressions for  $W,\Psi_a$ \cite{Cai-1-16}:
\begin{gather}
W=\frac 12(\rho_{\mathrm{sc}}-P_{\mathrm{sc}}), \\
\Psi_{a}=-\frac 12(\rho_{\mathrm{sc}}+P_{\mathrm{sc}})H\tilde{r
}dt+\frac12(\rho_{\mathrm{sc}}+P_{\mathrm{sc}})adr , \label{w psi}
\end{gather}
In those equations, we have considered the quantal correction.

The energy thrill through the apparent horizon at the interval time
$dt$ is given as \cite{Cai-1-16}
\begin{equation}
\delta Q=-dE=-A\Psi=A(\rho_{\mathrm{sc}}+P_{\mathrm{sc}})H
\tilde{r}_{\mathrm{A}}dt ,  \label{Q}
\end{equation}
with the area of the apparent horizon
$A=4\pi\tilde{r}_{\mathrm{A}}^2$. If the modified entropy of the
apparent horizon is denoted as $S_{\mathrm{A}}^{ \mathrm{sc}}$ ,
considering $\delta Q=TdS_{\mathrm{A}}^{\mathrm{sc}}$ and
Eq.(\ref{Q}), one can obtain
\begin{eqnarray}
dS_{\mathrm{A}}^{\mathrm{sc}}&=&8\pi^2\frac{1}{H^3}(\rho_{\mathrm{sc}}+P_{
\mathrm{sc}})dt\nonumber\\
&&=\frac{1}{
8\pi G}\left[\frac{2\pi\dot{A}}{\mathcal{S}}+\frac{A^2H^2}2
\frac{\dot{\mathcal{S}}}{\mathcal{S}^2}\right.\nonumber\\
&&\left.-\frac12\frac{\dot{\mathcal{S}}A^2}{a^2}
-\frac{1-\mathcal{S}^2}{a^2}\frac{HA^2}{\mathcal{S}}\right]dt\label{dS}
\end{eqnarray}
in which we have used the temperature of the apparent horizon
$T_{\rm A}=1/(2\pi \tilde{r}_{\mathrm{A}})$ with the radius of the
apparent horizon $\tilde{r} _{\mathrm{A}}=1/\sqrt{H^2+1/a^2}$ and Eq.(\ref{dotH}).
Expanding the above equation and integrating it, Eq.(\ref{dS}) can be written as
\begin{eqnarray}
S_{\mathrm{A}}^{\mathrm{sc}}&=&S_{\mathrm{Aa}}^{\mathrm{sc}}+\frac{A_{\mathrm{eff}}}{4G}+\frac{1}{4G}\int A_{\rm{eff}} d\ln\mathcal{S}\nonumber\\
&&+\frac{1}{8\pi G}\int\frac{A_{\rm{eff}}^2H^2}2
d\mathcal{S}-\frac{1}{16\pi G}\int \frac{A^2_{\rm{eff}}}{a^2\mathcal{S}^2} d\mathcal{S}\nonumber\\
&&-\frac{1}{8\pi G}\int \frac{(1-\mathcal{S}^2)\mathcal{S}}{a^2}{HA^2_{\rm{eff}}}dt,  \label{dseff}
\end{eqnarray}
in which $A_{\mathrm{eff}}=A/\mathcal{S}$ denotes the effective area
of the apparent horizon in the semiclassical LQC.
in which $S_{\mathrm{Aa}}^{\mathrm{sc}}$ is a constant and the value depends
on the specific physics.

In a short conclusion, considering the Einstein equation
(\ref{EinEq}) and assuming that the apparent horizon has Hawking
temperature, we get the expression for the modified  entropy-area
relation in the semiclassical LQC. The semiclassical theory not only
modify the entropy of apparent horizon, but also  the area of it.
And it is worth to notice that this modified entropy-area  relation
is not as same as the correction of black hole in loop  quantum
gravity \cite{Kaul-3-9}
\begin{equation}
S_{\mathrm{bh}}=S_{\mathrm{BH}}+\underset{n=0}{\overset{\infty}{\sum}}%
C_nA_H^{-n} ,  \label{Sbh}
\end{equation}
with some  constants $C_n$.  And it is different from the modified
entropy-area  relationship, which has the same form as
Eq.(\ref{Sbh}), when one considers the effective LQC with holonomy
corrections \cite{Cai-1-16}. This difference is reasonable. Because
we consider the inverse volume correction in the semiclassical
regions, but the  authors of \cite{Cai-1-16} considered the holonomy
correction. Notice that there are many differences between inverse
volume correction and holonomy correction. For example, the inverse
volume correction modifies the Klein-Gorden equation
\cite{Shinji-3-10}, but holonomy correction does not
\cite{Artymowski-3-10}; and the semiclassical energy density and
effective pressure which caused by the inverse volume correction are
different from the ones caused by the holonomy correction
\cite{Coleland-3-10}.

\subsection{The thermodynamical variables of dark energy fluid}

In this section, we will discuss the thermodynamics properties of
the dark energy in the inner side of the apparent horizon. Some
authors advised that the temperature of dark energy fluid has the
same temperature of the apparent horizon which the temperature is
described by Hawking temperature, but this may be get a ill-defined
entropy when $\omega_{\mathrm{cl}}<-1$ \cite{Gong-1-24}. And other
authors find the dark energy either have negative entropy or
temperature in the classical regions
\cite{Gonzalez-1-20,Myung-1-21,Lima-1-22}. In this subsection, we
will consider the temperature and entropy of the dark energy in the
inner side of the apparent horizon. And we will assume the dark
energy is a ideal fluid and with the chemical potential. This
subsection, we will follow the method of \cite{Emmanuel-1-23} to
discuss the temperature and entropy of the dark energy in
semiclassical LQC scenario.

The entropy fluid and particle fluid are described by $S^{\mu}_{\mathrm{sc}}=s_{\mathrm{sc}}u^{\mu}$ and $N^{\mu}=nu^{\mu}$, with the particle
number density $n$ and the entropy density $s_{\rm sc}$, respectively. Considering they satisfy $%
S^{\mu}_{;\mu}=0, N^{\mu}_{;\mu}=0$, it is easy to obtain
\begin{eqnarray}
\dot{s}_{\mathrm{sc}}+3Hs_{\mathrm{sc}}=0 ,  \label{s-con} \\
\dot{n}+3Hn=0 .  \label{n-con}
\end{eqnarray}
The solutions for above equations are
\begin{equation}
n(a)=n_0\left(\frac{a}{a_0}\right)^3,\qquad
s_{\mathrm{sc}}(a)=s_0\left(\frac{a}{a_0}\right)^3.  \label{a}
\end{equation}

Now we consider the thermodynamics. We assume the thermodynamic
properties is described by the particle number density $n$ and the
temperature $T_{ \mathrm{sc}}$. We have assumed the quantum effect
will modify the temperature. This is reasonable. If assuming that
the modified energy density and the temperature is still satisfy the
Stenfan-Boltzmann law $\rho_{\mathrm{sc} }\propto
f(T_{\mathrm{sc}})$, the temperature $T_{\mathrm{sc}}$ should be
dependent on $a$, for the energy density is the function of the
scale factor $a$, and then the temperature should be modified by the
semiclassical quantum effect. We will see that the Stenfan-Boltzmann
law still is tenable in our model later on. We consider the Gibbs
law, $T_{\mathrm{sc}}\left(\frac{\partial P_{\mathrm{sc}}}{\partial
T_{
\mathrm{sc}}}\right)_n=\rho_{\mathrm{sc}}+P_{\mathrm{sc}}-n\left[\frac{
\partial\rho_{\mathrm{sc}}(a)}{\partial n} \right]_{T_{\mathrm{sc}}} $, then
the prefect fluid can be described by
\begin{equation}
T_{\mathrm{sc}}(a)\left[\frac{\partial
\omega_{\mathrm{sc}}(a)\rho_{\mathrm{ sc}}}{\partial
T_{\mathrm{sc}}(a)} \right]_n=(\omega_{\mathrm{sc}
}(a)+1)\rho_{\mathrm{sc}}-n\left[\frac{\partial\rho_{\mathrm{sc}}(a)}{
\partial n} \right]_{T_{\mathrm{sc}}}.  \label{Gibbs}
\end{equation}
Considering $\dot{\rho}_{\mathrm{sc}}(a)=\dot{n}\left(\frac{\partial
\rho_{ \mathrm{sc}}(a)}{\partial
n}\right)_{T_{\mathrm{sc}}(a)}+{\dot{T}_{\mathrm{sc}}(a)}
\left(\frac{\partial \rho_{\mathrm{sc}}(a)}{\partial
T_{\mathrm{sc}}(a)} \right)_n$ and Eqs.(\ref{Mod-Con-E}) and
(\ref{Gibbs}), we can get the evolution of modified temperature with
respect to the semiclassical EoS parameters
\begin{equation}
\left[\frac{3}{a}T_{\mathrm{sc}}(a)\omega_{\mathrm{sc}}(a)+T_{\mathrm{sc}
}^{\prime }(a) \right](\omega_{\mathrm{sc}}(a)+1)=T_{\mathrm{sc}
}(a)\omega^{\prime }_{\mathrm{sc}}(a).  \label{T}
\end{equation}
Just as the mention in Sec.\ref{sec2}, the modified EoS parameters $\omega_{
\mathrm{sc}}(a)$ are possible -1 crossing. It is obvious that the modified
temperature which is described by above equation is zero when $\omega_{
\mathrm{sc}}(a)=-1$ for $\omega^{\prime }_{\mathrm{sc}}(a)\neq 0$.
Integrating the above equation when $\omega_{\mathrm{sc}}(a)\neq-1$, we can
get the expression for $T_{\mathrm{sc}}(a)$
\begin{eqnarray}
T_{\mathrm{sc}}(a)&=&T_0\left(\frac{\omega_{\mathrm{sc}}(a)+1}{\omega_0+1}
\right)\left(\frac{1}{a^{3\omega_{\mathrm{sc}}(a)}} \right)  \notag \\
&&\times\exp\left[-3\int^1_a da\omega^{\prime }_{\mathrm{sc}}(a)\ln
a \right], \label{sT}
\end{eqnarray}
with the present temperature $T_0$. We have considered that the present
scale factor $a_0=1$. There are four points should be noticed. First, if $
\omega_0=-1$ and $T_0\neq0$, $T_{\mathrm{sc}}(a)$ will be divergent.
But if we consider $\omega_{\mathrm{sc}}(a)$ describes today's dark energy, $
\omega_{\mathrm{sc}}(a)$ should not be modified by the quantum effect for $
a_0\gg a_\star$ and is substituted by $\omega_{\mathrm{cl}}(a)$. In
this case, Eq.(\ref{T}) describes the temperature for the dark
energy in the classical regions. The temperature of dark energy with $\omega(a_0)=-1$ should be zero, for
$\omega^{\prime }_{\mathrm{cl}}(a)|_{a=a_0}\neq 0$ in Eq.(\ref{T}). So, $T_0=0$ when $\omega_0=-1$ if the EoS $\omega_{\mathrm{cl}}$ is a function of the scale factor.
 Second,
the temperature is zero when $\omega_{\mathrm{sc} }(a)=-1$. But this "-1" EoS
parameter is caused by the effective behavior. Third, for $ T_0$ and
$\omega_0+1$ have the same sign, if $\omega_{\mathrm{sc}}(a)>-1$ the
temperature is positive, the temperature is negative when
$\omega_{\mathrm{sc }}(a)<-1$, and $T_{\mathrm{sc}}(a)=0$ when
$\omega_{\mathrm{sc}}(a)=-1$, this is as same as the behavior of
temperature in the classical regions \cite{Emmanuel-1-23}. But
notice that $\omega_{\rm sc}$ are modified EoS parameters and
includes the quantum geometry effect. Fourth, the scale factor is in
the regions $a_i<a<a_\star$, so $\frac{1}{a}$ in Eq.(\ref{sT}) will
not divergent, this is different from \cite{Emmanuel-1-23} in which
the authors considered the thermodynamical properties of dark energy
in the classical regions.

Now we have the expressions for modified temperature and modified
energy density. It is easy to obtain the generalized
Stefan-Boltzmann law
\begin{eqnarray}
\rho_{\mathrm{sc}}(a)&=&\rho_0\left[\frac{T_{\mathrm{sc}}(a)}{T_0}\frac{
\omega_0+1}{\omega_{\mathrm{sc}}(a)+1}
\right]^{\frac{\omega_{\mathrm{sc}
}(a)+1}{\omega_{\mathrm{sc}}(a)}}  \notag \\
&&\times\exp\left[\frac{3}{\omega_{\mathrm{sc}}(a)}\int^1_ada\omega^{\prime
}_{
\mathrm{sc}}(a)\ln a\right],  \notag \\
&& \hspace{3cm} \mathrm{for} \quad \omega_{\mathrm{sc}}(a)\neq 0;
\notag
\\
\rho_{\mathrm{sc}}(a)&=&\rho_0\left[\frac{T_{\mathrm{sc}}(a)}{T_0}\frac{
\omega_0+1}{\omega_{\mathrm{sc}}(a)+1}
\right]\left(\frac{1}{a}\right)^3,
\notag \\
&&\hspace{3cm} \mathrm{for} \quad \omega_{\mathrm{sc}}(a)=0.
\label{rho-T}
\end{eqnarray}
It is worth to notice that the above equations are still valid for
$\omega_{ \mathrm{sc}}(a)=-1$, for $T_{\mathrm{sc}}(a)=0$ in this
moment.

Now we turn to consider the chemical potential and entropy. Consider the
Euler's relation of $T_{\mathrm{sc}}(a)s_{\mathrm{sc}}(a)=[1+\omega_{\mathrm{%
sc}}(a)]\rho_{\mathrm{sc}}(a)-\mu_{\mathrm{sc}}(a) n$, with the
chemical potential $\mu_{\mathrm{sc}}(a)$, the chemical potential
can be given as \cite {Emmanuel-1-23}
\begin{equation}
\mu_{\mathrm{sc}}(a)=\frac{1}{n}[(1+\omega_{\mathrm{sc}}(a))\rho_{\mathrm{sc}%
}(a)-T_{\mathrm{sc}}(a)s_{\mathrm{sc}}(a)].  \label{mu}
\end{equation}
Using the Gibbs law (\ref{Gibbs}) and the modified temperature (\ref{sT}),
it is easy to get the modified chemical potential
\begin{eqnarray}
\mu_{\mathrm{sc}}(a)&=&\mu_0\left[\frac{\omega_{\mathrm{sc}}(a)+1}{\omega_0+1%
} \right]\left[\frac{1}{a^{3\omega_{\mathrm{sc}}(a)}} \right]  \notag \\
&&\times\exp\left[-3\int^1_a da\omega^{\prime }_{\mathrm{sc}}(a)\ln a \right]%
  \label{mus},
\end{eqnarray}
with chemical potential at present $\mu_0=\frac{1}{n_0}[\rho_0(%
\omega_0+1)-T_0s_0]$. As before, the above equation is
well-defined when $\omega_0=-1$, for $T_0=0$. But the sign of $\mu_{\mathrm{%
sc}}(a)$ is independent on $\omega_{\mathrm{sc}}(a)$, for the sign of $\mu_0$
can be general. And the chemical potential can be wrote as a function of $T_{%
\mathrm{sc}}(a)$,
\begin{equation}
\mu_{\mathrm{sc}}(a)=\mu_0\frac{ T_{\mathrm{sc}}(a)}{T_0} .  \label{mu-T}
\end{equation}

It is easy to get the entropy of the dark energy fluid by considering the
Eq.(\ref{mu}). Remembering $S(a)=s(a)V(a)=s(a)a^3$, and using Eqs.(\ref{a}),%
(\ref{rho-T}) and (\ref{mu-T}), we can get the expression for
entropy.
\begin{eqnarray}
S_{\mathrm{sc}}(a)&=&s_0V(a)\left[\frac{T_{\mathrm{sc}}(a)}{T_0}\frac{%
\omega_0+1}{\omega_{\mathrm{sc}}(a)+1} \right]^{\frac{1}{\omega_{\mathrm{sc}%
}(a)}}  \notag \\
&&\times\exp\left[\frac{3}{\omega_{\mathrm{sc}}(a)}\int^1_ada\omega^{\prime }_{%
\mathrm{sc}}(a)\ln a\right],
\end{eqnarray}
in which we have considered the present scale factor $a_0=1$. When $\omega_{%
\mathrm{sc}}(a)=0$, combining Eq.(\ref{rho-T}) and Eq.(\ref{mu-T}), we find
it is a trivial result. Note that, above equation is regular at $\omega_{%
\mathrm{sc}}(a)=-1$, because if $\omega_{\mathrm{sc}}(a)=-1$, $T_{\mathrm{sc}%
}=0$. And if $\omega_0=-1$, then $T_0=0$, so the above equation is
still well-defined. It is easy to get
$s_0V_0=s_{\mathrm{sc}}(a)V(a)$.

In a short conclusion, in this subsection, we has discussed the
thermodynamics properties of dark energy in the inside of apparent
horizon, and get the expressions for temperature, chemical potential
and entropy. We find that those variables are corrected due to the
semiclassical quantum effect, and those corrections are connected
with the modified EoS parameters. Our method is following
\cite{Emmanuel-1-23}, in which the temperature is regular for the
vacuum when they discuss the thermodynamics properties in the
classical regions but the temperature will be divergent at $a=0$.
But due to the modification of semiclassical quantum effect, $a=0$
doesn't lie in the regions we interested, so the expression for
temperature always holds for a varying classical EoS
$\omega_{\mathrm{cl}}$. The expressions for chemical potential and
entropy are obtain in this subsection.

\section{Conclusion and discussion}

\label{sec4} In this paper, we have discussed the thermodynamics
properties of the dark energy in the semiclassical LQC scenario. The
common feature of the all models of dark energy is the EoS
parameters $\omega$ of dark energy fluid is included by the field
evolution. So, at the first, we consider the correction of EoS
parameters in the regions we interested. We find that the corrected
EoS parameters depends on the changing rate of
inverse volume factor $D_l$ and classical EoS parameters $\omega_{\mathrm{cl}%
}(a)$.

We discussed the area-entropy relation of the apparent horizon of
universe which is full of dark energy fluid. We found that the
effective LQC not only modify the entropy of apparent horizon, but
also the area of it. And the modified area-entropy relation is
different from the one considered by \cite{Cai-1-16}, in which the
authors considered the holonomy correction. And then, we discussed
the thermodynamical properties of dark energy fluid in the inner
side of apparent horizon. The expressions for temperature, chemical
potential and entropy are obtained. We find that, if the classical
EoS parameter is a function of the scale factor, the thermodynamical
variables are functions of the modified EoS parameters and
well-defined when $\omega(a_0)=-1$.  The sign of temperature is dependent on the modified EoS $\omega_{\mathrm{sc}}$, when $\omega_{\mathrm{sc}}>-1$,
$T_{\mathrm{sc}}>0$, $T_{\mathrm{sc}}<0$ when $\omega_{\mathrm{sc}}<-1$,  and $T_{\mathrm{sc}}=0$ if $\omega_{\mathrm{sc}}=-1$, but the sign of
the modified temperature is independent on the sign of potential, which is determined by the sign of $\mu_0$. This is as same as the discussion in the classical region \cite{Emmanuel-1-23}.  $T_{\rm
sc}(a)=S_{\rm sc}(a)=0$ when $\omega_{\rm sc}(a)=-1$, this is like
the behavior of massless degenerate Fermi
gas \cite{Bilic-4-2,Palash-4-3}. Just as we discussed above, when
$\omega_{\rm sc}<-1$, the temperature is negative. But this negative temperature is independent on the sign of potential, for
the sign of potential is just dependent on $\mu_0$.
The negative
temperature can only be explain in the quantum framework. The
quantum field theory for the field with EoS parameters $\omega_{\rm
cl}<-1$ is still an open question. In the semiclassical regions, the
EoS parameters should be modified by the quantum geometry effect, so
the negative temperature in the semiclassical regions partial is
cased by the quantum geometry effect.

In this paper, we just consider the thermodynamics properties of
dark energy fluid, it is easy to verify that our method can apply to the other kinds of
matter, e.g., stiff matter, radiation and so on, for the EoS of this matter will be modified in the effective region in loop quantum cosmology, just as Singh shown \cite{Singh-2-2}. But this model can not describe the thermodynamics properties of cosmological constant with the constant EoS -1. Although the EoS of cosmological constant will be modified in the effective region, just as Eq.(\ref{os-oc}) described. The temperature of cosmological constant today is still unclear, so the term $T_0/(\omega_0+1)$ in Eq.(\ref{sT}) will be divergent for $\omega_{\mathrm{sc}}$ does not always equal to -1.

\acknowledgments The work was supported by the National Natural
Science of China (No. 10875012) and the Scientific Research
Foundation of Beijing Normal University.


\begin{thebibliography}{99}
\bibitem{Alessandro-1-0} A. Melchiorri, L. Mersini, C. J.
Odman and M. Trodden, Phys. Rev. D \textbf{68}, 043509(2003).

\bibitem{Edmund-1-1} E. J. Copeland, M. Sami and S. Tsujikawa,
Int.J.Mod.Phys.D \textbf{15},1753(2006); R. Rakhi and K. Indulekha,
arXiv:0911.2601; A. Kesavan, \emph{Dark energy}, arXiv:0908.2852.

\bibitem{Guo-1-2} Z. K. Guo, N. Ohta and Y. Z. Zhang,
Phys. Rev. D \textbf{72}, 023504(2005); 

\bibitem{Singh-1-3} P. Singh, M. Sami and N. Dadhich,
Phys. Rev. D \textbf{68}, 023522(2003);
V. B. Johri, 
Phys. Rev. D \textbf{70}, 041303(R)(2004);
L. P. Chimento and R. Lazkoz, 
hep-th/0405518.

\bibitem{Fang-1-4} B. Fang, X. L. and X. M. Zhang, Phys. Lett. B \textbf{607},
35,(2005); W. Zhao and Y. Zhang, Phys. Rev. D
\textbf{73},123509(2006);
Y. P. Zhang, Z. L. Yi, T.J. Zhang and W. B. Liu, 
Phys. Rev. D \textbf{77}, 023502 (2008).

\bibitem{Qaisar-1-5} Q. Shafi, A. Sil and Siew-Phang Ng,
Phys. Lett. B \textbf{620}, 105(2005); A. de la Macorra, F.
Briscese,
AIP Conf. Proc. \textbf{1116}, 179(2009);
N. Bose, A. S. Majumdar,%
Phys. Rev. D \textbf{79}, 103517(2009);
Philippe Brax, Carsten van de Bruck, L. M. H. Hall, J. M. Weller,%
Phys. Rev. D \textbf{79}, 103508(2009).

\bibitem{Guo-1-6} Z. K. Guo, Y. S. Piao, and Y. Z. Zhang, 
Phys. Lett. B \textbf{594},  247(2004); 
J. G. Hao and X. Z. Li,
Phys. Rev. D \textbf{68}, 083514(2003).


\bibitem{Maia-1-8} M. D. Maia, E. M. Monte, J. M. F. Maia, and J.S.
Alcaniz, 
Class. Quant. Grav. \textbf{22},  1623(2005); M. D. Maia, A. J. S.
Capistrano, J. S. Alcaniz, and E. M. Monte,
arXiv:0905.4259.

\bibitem{Sami-1-7} M. Sami,
Mod. Phys. Lett. A \textbf{19}, 1509 (2004); R. R. Caldwell, M.
Kamionkowshi, and N. N. Weinberg,
Phys. Rev. Lett. \textbf{91}, 071301(2003).


\bibitem{Nojiri-1-9} S. Nojiri, S. D. Odintsov, and  S.
Tsujikawa,
Phys. Rev. D \textbf{71}, 063004(2005); S. Carneiro, R. Tavakol,
Gen. Rel. Grav. \textbf{41}, 2287(2009).

\bibitem{Pereira-1-11} P. F. Gonzalez-Diaz, and C. L. Siguenza,
Nucl. Phys. B \textbf{697}, 363(2004);
S. H. Pereira and J. A. S. Lima, 
Phys. Lett. B \textbf{669}, 266(2008).
\bibitem{Singh-1-10} J.S. Alcaniz, D. Jain, A. Dev,
Phys. Rev. D \textbf{66}, 067301(2002); M. Sami, P. Singh, S.
Tsujikawa,
Phys. Rev. D \textbf{74}, 043514(2006); S.B.Chen, B. Wang, J.L.
Jing,
Phys. Rev. D\textbf{78} 123503(2008); X. Y. Fu, H. W. Yu,and P. X.
Wu,
Phys. Rev. D\textbf{78}, 063001(2008);
 A. de la Macorra, F.
Briscese,
AIP Conf. Proc. \textbf{1116}, 179(2009); E. S. Corchero,
AIP Conf. Proc.\textbf{1122}, 229(2009); Y.P. Zhang, Z.L. Yi, T.J.
Zhang, and W.B. Liu,
Phys. Rev. D \textbf{77},023502(2008).

\bibitem{Hawking-1-13} G. W. Gibbons and S. W. Hawking,
Phys. Rev. D \textbf{15}, 2738(1977); M. D. Pollock and T. P. Singh,
Class. Quant. Grav. \textbf{6}, 901(1989);
A. V. Frolov and L. Kofman, 
JCAP \textbf{0305},009(2003).

\bibitem{Wang-1-14} B. Wang, Y. G. Gong and E. Abdalla,
Phys. Rev. D \textbf{74}, 083520 (2006).

\bibitem{Li-1-15} L. F. Li and J. Y. Zhu,
Advances in High Energy Physics, 2009, 905705(2009)

\bibitem{Cai-1-16} R. G. Cai, L.M . Cao and Y. P. Hu,
JHEP \textbf{08}, 090(2008).

\bibitem{Bojowald-1-17} M. Bojowald, 
Living Rev. Rel. \textbf{11}, 4(2008).

\bibitem{Ashtekar-1-18} A. Ashtekar,
J. Phys .Conf. Ser. \textbf{189}, 012003(2009);
A. Ashtekar, 
Gen. Rel. Grav. \textbf{41}, 707(2009).

\bibitem{Ashtekar-1-19} A. Ashtekar, M. Bojowald and J.
Lewandowski, 
Adv. Theor. Math. Phys. \textbf{7}, 233(2003).

\bibitem{Emmanuel-1-23} E. N. Saridakrics, P. F. Gonzalez-Diza and
C. L. Siguenza, 
Class. Quantum Grav. \textbf{26}, 165003(2009).

\bibitem{Gonzalez-1-20} P. F. Gonzalez-Diaz and C. L. Siguenza,
Nucl. Phys. B \textbf{697}, 363(2004).

\bibitem{Myung-1-21} Y. S. Myung,
Phys. Lett. B \textbf{671}, 216(2009).

\bibitem{Lima-1-22} J. A. S. Lima, J. S. Alcaniz,
Phys. Lett. B \textbf{600}, 191(2004); J. A. S. Lima, S. H. Pereira,
Phys.Rev.D \textbf{78}, 083504(2008).

\bibitem{Gong-1-24} Y. G. Gong, B. Wang and A. Z. Wang,
JCAP \textbf{01}, 024(2007); Y. G. Gong, B. Wang and A. Z. Wang,
Phys. Rev. D \textbf{75}, 123516(2007)

\bibitem{Singh-2-1} J. Magueijo and P. Singh,
Phys. Rev. D\textbf{76}, 023510(2007).

\bibitem{Vandersloot-2-1} K. Vandersloot,
Phys. Rev. D\textbf{71}, 103506(2005).

\bibitem{Singh-2-2} P. Singh,
Class. Quantum Grav. \textbf{22}, 4203(2005).


\bibitem{Frolov-3-1} A. V. Frolov and L. Kofman,
JCAP \textbf{05}, 009(2003).

\bibitem{Danielsson-3-2} U. H. Danielsson,
Phys. Rev. D \textbf{71}, 023516(2005).

\bibitem{Bousso-3-3} R. Bousso,
Phys. Rev. D \textbf{71}, 064024(2005).

\bibitem{Poisson-3-5} E. Poisson and W. Israel,
Phys. Rev. D \textbf{41}, 1796(1990).

\bibitem{Misner-3-6} C. M. Misner and D. H. Sharpe,
Phys. Rev.\textbf{136}, B571(1964).


\bibitem{Kaul-3-9} R. K. Kaul and P. Majumdar,
Phys. Rev. Lett. 84, 5255(2000).

\bibitem{Shinji-3-10}S. Tsujikawa, P. Singh and R. Maartens, 
Class. Quan. Grav. \textbf{21}, 5767, (2004).

\bibitem{Artymowski-3-10}M. Artymowski, Z. Lalak and L. Szulc, 
JCAP \textbf{01}, 04(2009).

\bibitem{Coleland-3-10}E.J. Copeland, D. J. Mulryne, N. J. Nunes and
M. Shaeri, 
Phys. Rev. D \textbf{77}, 023510(2008).

\bibitem{Bilic-4-2}N. Bilic, 
Fortschr. Phys. \textbf{56}, 363(2008).

\bibitem{Palash-4-3} P. B. Pal, \emph{An Introductory Course of
Statistical Mechanics} (Alpha Science, Oxford, U.K.).
\end{thebibliography}
\end{document}